\documentclass[manuscript]{aastex}
\usepackage{graphics}
\usepackage{mathptmx}
\usepackage[usenames,dvipsnames]{color}

\def\kms{\mbox{${\rm km}\:{\rm s}^{-1}\,$}}

\def\mloss{\mbox{$M_\odot \: yr^{-1}\:$}}

\shorttitle{The evolutionary status of UIT\,005}
\shortauthors{Urbaneja et al.}

\begin{document}

\title{Blue luminous stars in nearby galaxies--UIT\,005: a possible link to the Luminous Blue Variable stage 
\footnote{Based on observations made with the William Herschel Telescope operated on the island of 
La Palma by the Isaac Newton Group in the Spanish Observatorio del Roque de los Muchachos of the 
Instituto de Astrof\'{\i}sica de Canarias.} }

\author{M. A. Urbaneja}
\affil{Institute for Astronomy, University of Hawaii, 2680 Woodlawn Drive,
     Honolulu, Hawaii 96822}

\author{A. Herrero} 
\affil{Instituto de Astrof\'{\i}sica de Canarias,
    V\'{\i}a L\'actea S/N, E-38200 La Laguna, Canary Islands, Spain.\\
Dpto. de Astrof\'{\i}sica, Universidad de La Laguna, Avda.
Astrof\'{\i}sico Francisco S\'anchez, E-38205 La Laguna, Canary Islands, Spain}

\author{D. J. Lennon}
\affil{ESA, Space Telescope Science Institute, 3700 San Martin Drive, Baltimore, MD 21218}

\author{L. J. Corral}
\affil{Instituto de Astronom\'{\i}a y Metereolog\'{\i}a, Universidad de Guadalajara, Avda. Vallarta 2602,
Guadalajara, Jalisco, C.P. 44130, Mexico}

\and

\author{G. Meynet}
\affil{Geneva Observatory, Ch. des Mailettes 51, 1290 Sauverny, Switzerland}

\begin{abstract}
A detailed study of the blue supergiant UIT\,005 (B2-2.5Ia$^+$) in M\,33 is presented. The results of our quantitative spectral 
analysis indicate that the star is a very luminous, $\log \mathrm{L}/\mathrm{L}_\odot\sim5.9\,\mathrm{dex}$, and 
massive, M$\sim$50\,M$_\odot$, object, showing a very high nitrogen-to-oxygen ratio in its surface (N/O$\sim$8, by mass).
Based on the derived Mg and Si abundances, we argue that this high N/O ratio cannot be the 
result of an initial low O content due to its location on the disk of M\,33, a galaxy known to present 
a steep metallicity gradient. In combination with the He abundance, the most plausible interpretation is that UIT\,005
is in an advanced stage of evolution, showing in its surface N enrichment and O depletion resulting from mixing 
with CNO processed material from the stellar interior. 
A comparison with the predictions of current stellar evolutionary models indicates that there
are significant discrepancies, in particular with regard to the degree of
chemical processing, with the models predicting a much lower degree of O depletion than observed. 
At the same time, the mass-loss rate derived in our analysis is an order of magnitude lower 
than the values considered in the evolutionary calculations.
Based on a study of the surrounding stellar population and the nearby
cluster, NGC588, using HST/WFPC2 photometry, we suggest that
UIT\,005 could be in fact a runaway star from this cluster. Regardless of its origin, the derived parameters place the star in a region of the Hertzsprung--Russell diagram where 
Luminous Blue Variables are usually found, but we find no evidence supporting photometric or 
spectroscopic variability, except for small H$\alpha$ changes, otherwise observed in Galactic B-type 
supergiants. Whether UIT\,005 is an LBV in a dormant state or a
regular blue supergiant could not be discerned in this study. Subsequent monitoring would help us to improve
our knowledge of the more massive stars, bridging the gap between regular and more exotic blue 
supergiants. 

\end{abstract}

 %%, winds, outflows}

\section{Introduction  \label{section_intro} }

There is no doubt that massive stars, those born with masses above $\sim$8\,M$_\odot$, played, and 
still play, a 
fundamental role in shaping the universe as we observe it. Unmistakable signatures of their presence are 
imprinted in the spectra of star forming galaxies, both nearby and at high redshift \citep{steidel1996}. It is 
in fact because of their very existence that the physical properties of these distant galaxies can be 
investigated, 
through directly related stellar indicators such as stellar wind features
\citep{rix2004}, or indirectly related stellar signatures, such as the
ionizing radiation field reprocessed by the gas and radiated in a few 
forbidden emission lines \citep{kobulnicky2000, pettini2001}. Throughout their entire life, from the Main
Sequence to their 
end in a supernova event, massive stars drive, to a significant extent, the evolution of galaxies via 
radiative, kinetic and chemical feedback. Understanding their physical properties and their evolution in different 
environments is therefore a relevant issue for many different fields in modern Astrophysics. \\

A proper understanding of these objects needs to be assessed by testing the level of 
agreement between the different theories involved, namely the atmosphere and interior
structures. Predictions under different conditions have to be compared at any given 
evolutionary stage. This has not been a trivial task for
the last two decades (see \citealt{maeder2000}, or the review by \citealt{herrero2008}), and although 
in the last years an increasingly better agreement has 
been reached, there are still some pending issues, such as the efficiency of rotational mixing 
or the presence of other phenomena affecting the stellar evolution \citep{trundle2005, hunter2008, maeder2009}.

Particularly important in this context is the role that stars in nearby galaxies can play, where very 
different conditions can be found.  
Only a relatively small number of Galactic massive stars are available for these comparative
studies, since the massive stellar population is basically concentrated in the Galactic plane,
where the thick clouds of dust keep them hidden. This small number makes the
observation of short-lived evolutionary phases, which could be crucial as they
are affected by strong episodes of mass-loss and mixing \citep{massey2003},
even more difficult. The problem is less severe in nearby galaxies, such as M\,33, where we 
can have access to a much larger number of massive stars and objects in these short-lived phases.\\

We came across a very interesting object in our spectroscopic survey
of massive blue supergiants in M\,33. It seemed quite probable that UIT\,005 
(RA: 01$^h$32$^m$42.89$^s$,  Dec: +30$^{\circ}$38$^\prime$47.2$^{\prime\prime}$, J2000, \citealt{massey1996})
could be in an advanced phase of evolution, one of those very short-lived phases
and close to the LBV stage. The aim of this paper is to test the level of agreement between the results obtained from the model atmospheres 
and the theory of stellar evolution in such object. This will provide us with clues 
about the consistency between these models in an environment and evolutionary stage 
different from the most studied cases. 

The rest of this paper is organized as follows; we describe in \S\ref{obs} the observations and data reduction steps. The different aspects of the spectral 
analysis are presented in \S\ref{spec} and the results are discussed in \S\ref{disc}. Finally, our 
conclusions are presented in \S\ref{conc}. Throughout this work, metal abundances are expressed as 
A$\left(\mathrm{X}\right)\,=\,\log\left(N\left(\mathrm{X}\right)/N\left(\mathrm{H}\right)\right)+12$. Regarding the host galaxy, 
we adopt a distance modulus of $\mu$\,=\,24.56$\,\pm\,$0.1\,mag \citep[][]{freedman2001},  and 
R$_{25}\,=\,$28.77$^\prime$,  a position angle of the line of nodes $\phi$\,=\,22$^{\circ}$\,and inclination $i$\,=\,54$^{\circ}$~\citep{vilacostas1992}, with the galactic center
located at RA\,01$^h$33$^m$50.9$^s$~Dec\,+30$^{\circ}$39$^\prime$36$^{\prime\prime}$ (J2000).

\section{Observations \label{obs}}

\subsection{Photometry \label{section_photometry}}
%% Note: Wide Field Camera 2 scale: 0.1 arcsec/pixel
%%
Photometric information for UIT\,005 is available in several
bands from diverse sources. Its designation as UIT\,005 originates from the work by \citet{massey1996}, 
where, along with fluxes in two ultraviolet bands, U-, B- and V-band photometry was provided. 
\citet{massey2006} published new  U-, B-, V-, R- and I-band photometric measurements in a recent 
epoch. In this work, UIT\,005 can be identified as the source J013242.9+303847.0. 
A much fainter source, J013242.86+303847.0, is almost coincident with our star (see below).

\begin{figure}[!]
   \begin{center}
   \includegraphics[width=0.45\textwidth]{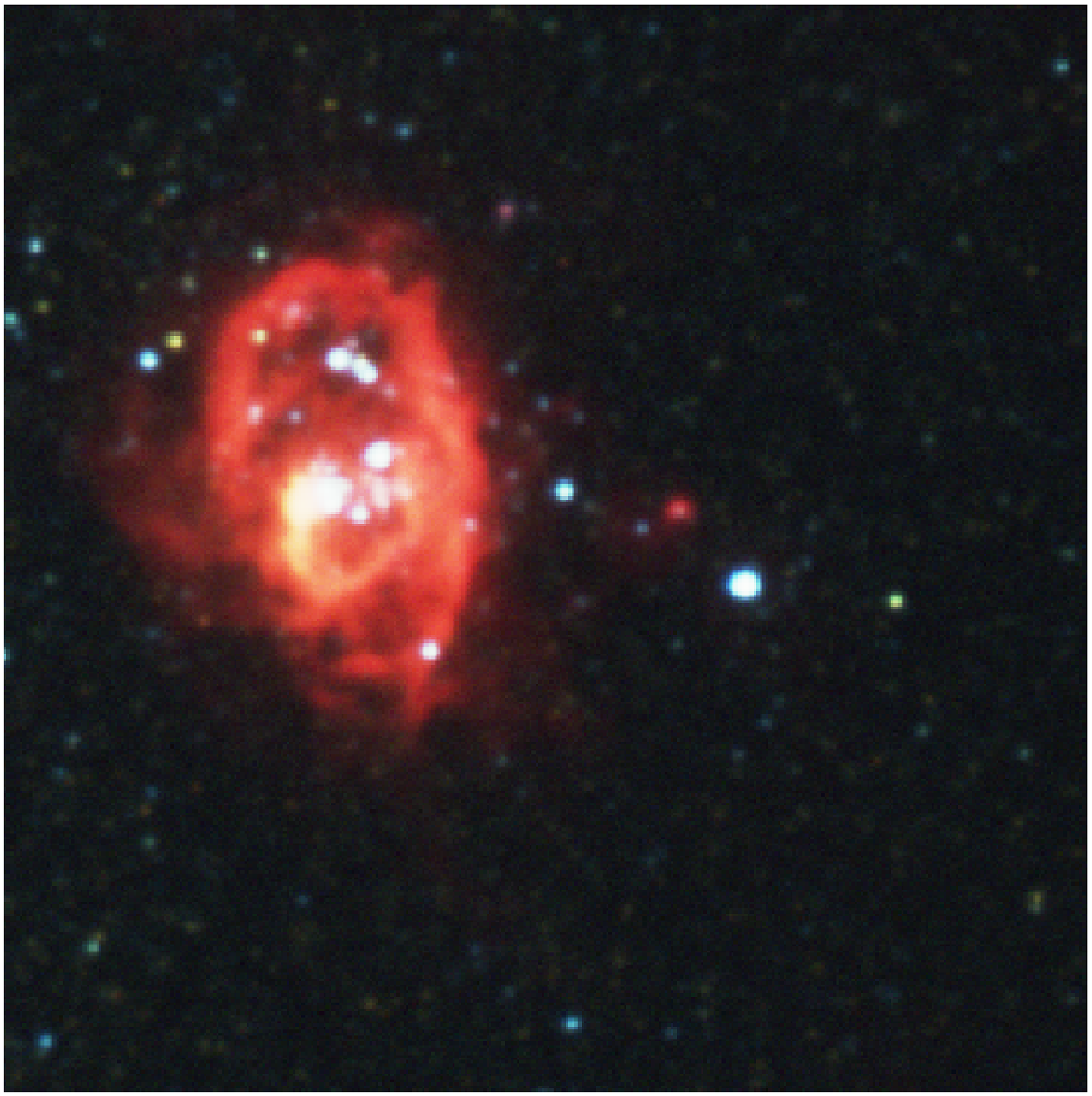}   
   \includegraphics[width=0.484\textwidth]{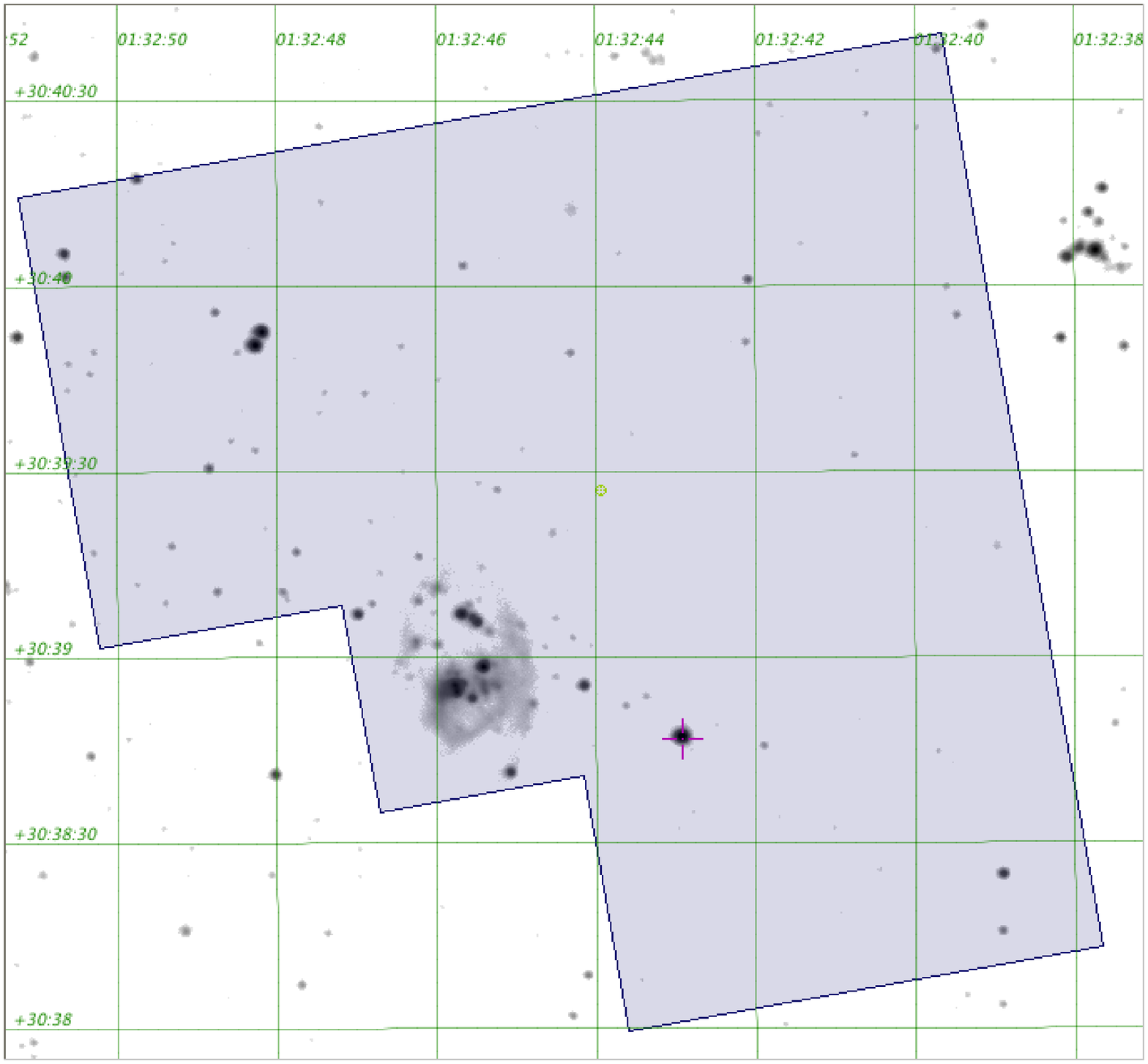}        
   \caption[ ]{(Left) False color composite image of a 1.5$\times$1.5 arcmin$^2$ field around UIT\,005. North is up and East is to the left; UIT\,005 is the 
   blue bright source slightly to the West of the image's center. The nearby giant \ion{H}{2} region NGC\,588 is included. This view results 
   from the combination of three images from the Local Group Galaxy Survey \citep{massey2006}: red--H$\alpha$, 
   green--V, and blue--B.  (Right) A wider field is shown with the WFPC2 footprint overlaid on a V band image; UIT\,005 is identified by a red cross. The orientation
   of both images is the same. 
   \label{fig_massey2006}}
   \end{center}
\end{figure}

The nearby giant \ion{H}{2} region NGC\,588 and its ionizing cluster were targeted with the 
Wide-Field and Planetary Camera 2 (WFPC2) on board of the Hubble Space Telescope (HST) during 
Cycle 4 (program 5384), using a suite of 5 different filters: F170W, F336W, F439W, F469W and F547M.
Given its close proximity (both objects are separated by $\sim$30$^{\prime\prime}$, see 
Fig.\,\ref{fig_massey2006}), UIT\,005 was imaged by one 
of the three detectors of the Wide-Field Camera. The fainter source mentioned before is spatially 
resolved in the HST images. We have carried out the photometry of these data using HSTphot 
\citep[][]{dolphin2000a}. Magnitudes in the WFPC2 filter system can be transformed to the 
Johnson-Cousins system by applying the recipes presented by \citet{holtzman1995} and 
\citet{dolphin2000b}.

The contamination of \citet{massey1996} photometry due to the faint source can be
estimated, either by using \citet{massey2006} or our WFPC2 photometry. Both
sets indicate a contribution to the combined flux of $\sim$2\%~for the  
U, B and V bands. This would not the case for the R-band where, according to
the values published by \citet{massey2006}, the faint source is only $\sim$0.2 mag 
fainter than UIT\,005, which translates into a flux ratio of 1:0.78. This is unexpected, 
given the almost constant flux ratio in the bluer bands, and the
fact that UIT\,005 was detected in the I-band, whilst the faint source was
not. Moreover, there is no indication of a second source in the red spectra (see
later); if the R-band magnitudes are right, then we would overestimate the
continuum by almost a factor 2, which would severely affect the measured
equivalent widths of the metal lines around H$\alpha$. We conclude that 
the published R-band magnitude of the faint source is affected by some
unknown problem, and that it is not representative of its real value.

If we disregard for a moment the possibility of photometric variability for
UIT\,005, and combine the WFPC2/HST photometry with the previous ground based values, we 
find for the mean values V\,=\,16.79$\,\pm\,$0.05, B-V\,=\,0.014$\,\pm\,$0.014 and 
U-B\,=\,-0.942$\,\pm\,$0.099 mag. The uncertainties associated with these values are an order 
of magnitude larger than the individual 1$\sigma$ uncertainties quoted in the 
different photometric measurements. The small scatter in the V-band 
magnitude and almost constant B-V color suggest a low level of photometric variability, within the 
expectations for blue supergiant stars. At the same time, the scatter in U-B is a little bit
higher, indicating perhaps photometric variations in the U-band.

The combination of the different photometric observations could be affected
in principle by the time elapsed between the measurements. Most
unfortunately, the possibility of time variability could not be investigated with any of the recent
surveys of M\,33; UIT\,005 is located in a region of the galaxy that was not monitored either 
by the DIRECT survey \citep{macri2001} nor by the more recent WISE observatory M\,33 variability project
\citep[][]{shporer2006}. With regard to the deep survey of the whole galaxy
introduced by \citet[][]{hartman2006}, UIT\,005 appears saturated in the images (D. Bersier, private communication).

Regarding longer wavelengths, UIT\,005 can be matched to the object  
2MASS\,01324287+3038468 from \citet{cutri2003}; the star was detected only
in the J-band. Full coverage in near-IR bands have been obtained from the IR survey presented by
\citet{cioni2009}; the agreement in the J-band magnitudes of these two surveys is very good. Our 
object is also cross-identified as the Spitzer Space Telescope source SSTM3307\,J01324289+303847.1 in the recent work 
by \citet[][]{mcquinn2007}. Infrared Array Camera (IRAC) magnitudes, obtained in five different epochs
spanning 19 months from January 2004 to January 2006, in the 3.6 and 4.5\,$\mu$m channels can be 
found in this last reference; UIT\,005 was not detected at 8.0\,$\mu$m. The values shown in 
Tab.\ref{table_phot} are the mean values of those five individual measurements. The uncertainties in the
individual measurements are comparable to the error in the mean values, which is consistent with the fact 
that \citet{mcquinn2007} did not consider UIT\,005 among the variable stars in their sample. \\

%% Also, archival MIPS 24\,$\mu$m images (see Fig.\,\ref{fig_spitzer}) do not show point source emission at the location of UIT\,005. Multiband Imaging Photometer, MIPS. \\ 
%comment-GALEX: whilst observed, the star in not included in the catalog. UIT FUV and NUV magnitudes are available from \citet{massey1996}, but they do not compare well with our model. Could this be related to a significant departure of the actual extinction curve from the one used to redden the synthetic SED?

The photometric measurements, along with the corresponding dates, are summarized in Tab.\,\ref{table_phot}. 
We show in Fig.\,\ref{fig_massey2006} a false color image of a 1.5$'\times$1.5$'$ field 
approximately centered on UIT\,005, using B (blue), V (green) and H$\alpha$ (red) images from the Local 
Group survey \citep{massey2006}.  From the low level variations in V and B-V, and the non-variability assigned
by \citet[][]{mcquinn2007}, it seems reasonable to conclude that the star is not showing photometric
variations beyond the stochastic level. The changes in the U-band would be most likely related to 
issues with the calibration of the different ground-based measurements. As it will be shown 
later (see \S\ref{disc}), all these photometric measurements, excluding the U-band, are well
reproduced by the spectral energy distribution (SED) predicted by the model corresponding to our derived 
fundamental parameters.

\subsection{Spectroscopy \label{section_spectroscopy}} 

The quantitative analysis presented in the following sections is mainly based on
two observational runs with the 4.2m William Herschel Telescope (WHT) equipped 
with the Intermediate dispersion System (ISIS). The blue wavelength range (3970--4850 \AA) was 
acquired in September 7th 2000, 
using a 1200 lines\,mm$^{-1}$ grating (dispersion, 0.23 \AA\,pix$^{-1}$) centered at 4400 \AA. 
The 1.2 arcsec slit was imaged with an EEV12 detector (scale, 0.19 arcsec\,pix$^{-1}$),
providing a spectral resolution of $\sim\,$0.9 \AA~ full width half maximum (FWHM). Two 
3600 s exposures were secured under a seeing of 2 arcsec, for an airmass below 1.3.

The red spectrum was observed in October 15th 2003, using ISIS red arm
equipped with a 600 lines\,mm$^{-1}$ grating (dispersion, 0.44 \AA\,pix$^{-1}$). A 1.2 arcsec slit was
used, imaged with the MARCONI2 chip (scale 0.20 arcsec\,pix$^{-1}$), providing the system a 
spectral resolution of 1.6 \AA~FWHM. Two exposures, one of 2200 s and another of
2000 s, were acquired, for a characteristic airmass below 1.4. The red spectrum expands 
approximately 1200 \AA, centered at 6558 \AA.\\

A third data set was collected from the Keck Observatory Archive. The star was observed
in October 10th, 2001 with the High Resolution Echelle Spectrometer \citep[HIRES,][]{vogt1994} 
at Keck\,1. This set contains 4 exposures of 3600 s each, covering    
 the range $\sim$4265--6630 \AA~at R$\sim$65000, 
with gaps in the coverage between some of the 30 orders. Unfortunately, the instrumental set-up used 
during these observations was not optimal for B-type stars, and a number of key diagnostic 
lines, in particular the \ion{Si}{2} and \ion{Si}{4} lines that are required to determine the effective 
temperature, were not included. Whilst
an independent analysis of this spectrum is then not possible, we can still compare it with the ISIS 
spectra, to search for spectral variations. \\

The ISIS data were processed using standard IRAF\footnote{IRAF is distributed by the National Optical 
Astronomy Observatories, which are operated by the Association of Universities for Research in 
Astronomy, Inc., under cooperative agreement with the National
Science Foundation.} routines for long slit spectroscopy (bias 
correction, flat-fielding, wavelength calibration and optimal extraction). The archival HIRES data were fully 
processed with Tom Barlow's MAKEE package. The final extracted spectra were continuum normalized using 
low-orders polynomial fits to well defined continuum points.\\

New, service mode, observations with ISIS/WHT were carried out on November 6th, 2009, under
clear sky conditions, for a seeing of $\sim$1.2 arcsec and a characteristic airmass of 1.7. In this 
occasion, both arms were fed simultaneously. The configuration of the blue arm was quite similar to the
one used in the 2000 run, but with the spectrum centered at 4260 \AA, hence covering the range 
3868--4618 \AA. For the red arm, we selected the R1200R grating, centered at 6800~\AA, with the spectrum
covering the range 6497--7190 \AA~at a spectral resolution of $\sim$1 \AA~FWHM for a projected slit width 
of 1.2 arcsec. Two exposures of 1800 s each were collected. This set was processed in the same way as
explained above. The S/N achieved (S/N$\sim$30-40) is well below the one in the 2000/2003
cases (S/N$\sim$80) ; nevertheless, this new spectrum shows that the star has not changed.\\    

%% NOTE : 2009 blue S/N 50@4100 - 30@4600 
%% NOTE : 2009 red S/N ~30
%% NOTE : 2000 blue S/N ~85
%% NOTE : 2003 red S/N ~70 in Halpha

We have carefully compared the spectra from the various
epochs for radial velocity variations, finding no
significant evidence for binarity.  However due to
difficulties in assigning zero-point off-sets for the
various observational configurations we are unable
to detect radial velocity variations smaller than
$\pm$10 \kms.

\section{Spectral Analysis\label{spec}}

\subsection{Qualitative description of the spectrum}

The first noteworthy spectral characteristic is the narrow H Balmer profiles, pointing to 
a high luminosity. This is supported by the 2003 H$\alpha$ profile, fully in emission. In terms of metal
features, the blue spectrum is characterized by the presence of strong \ion{N}{2} lines, and the
weakness of the \ion{O}{2} lines. Other prominent features present in the spectrum are the strong
\ion{Si}{3} triplet lines at 4553-67-74 \AA~and the \ion{Mg}{2} doublet at 4481 \AA. Three
other strong \ion{N}{2} lines appear in the observed ISIS red domain, along with H$\alpha$ and 
\ion{He}{1} 6678 \AA~line, plus the \ion{C}{2} doublet at 6578-82~\AA.   

\begin{figure}[!]
 \begin{center}
 \includegraphics[]{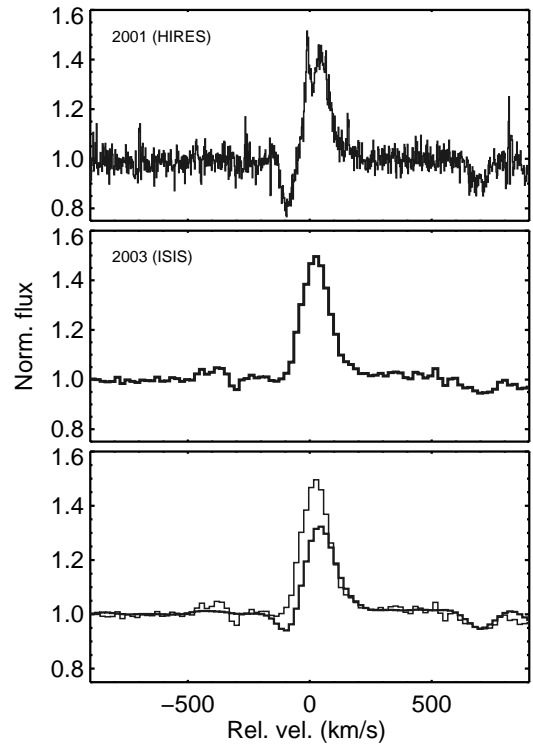} 
 \caption[ ]{UIT\,005 H$\alpha$ profiles. From top to bottom: HIRES (2001, R$\sim$65000), 
 ISIS (2003, R$\sim$5000), and a comparison of the two observations with the HIRES profile (thick) 
 degraded to the ISIS resolution (thin profile). \label{fig_halpha}}
 \end{center}
\end{figure}

The two H$\alpha$ profiles, shown in Fig.~\ref{fig_halpha}, are slightly different. In 2001, the profile 
(top panel in Fig.~\ref{fig_halpha}) appears as a P Cygni line, whilst the 2003 observation 
(mid panel) shows a strengthening of the profile in pure emission. To better illustrate this change, we have
included in the bottom panel of Fig.~\ref{fig_halpha} a comparison of both observations, with the HIRES 2001 
profile degraded to the spectral resolution of the ISIS 2003 observation. By contrast, the few metal lines 
in the the red range do not show any change, within the noise level. In the same sense, 
there are no noticeable differences in the blue range between the ISIS 2000 and HIRES 2001 
spectra.\\

By comparing the equivalent width ratios of \ion{Mg}{2} 4481 \AA~to the \ion{Si}{3} 
triplet lines with the sample of Galactic supergiant stars of \cite{lennon1993} 
and the LMC sample of \cite{fitzpatrick1991}, we tentatively 
locate the star in the spectral type range of B2-2.5. Although the metallicity 
of UIT\,005 could be different from that of these templates, the 
abundance ratio of these two elements remains fairly constant \citep[see][]{urbaneja2005}. 
Correspondingly, we assign a luminosity class of Ia$^+$ due to the narrow H Balmer lines. This
is a minor revision in luminosity class with respect to the classification previously provided by \cite{massey1996} on
the basis of low resolution spectroscopy (see their Fig~5b, displaying a 
spectrum covering the wavelength range $\sim$3900--4900 \AA, obtained either in 1993 or 1995).  

\subsection{Methodology: model atmosphere/line formation code and main assumptions}

The model atmosphere/line formation code {\sc fastwind} (Fast Analysis of STellar atmospheres
with WINDs) was used to create a small set of models for the analysis. The code was first introduced
by \cite{santolayarey1997}. Meanwhile, a number of improvements have been incorporated. A 
complete description of the current status of the code as well as comparisons with alternative  
codes have been presented by \cite{puls2005}. Briefly, as used in this work, {\sc fastwind} solves the 
radiative transfer problem in the co-moving frame of the
expanding atmospheres of early-type stars in a spherically symmetric geometry,
subject to the constraints of statistical equilibrium and energy 
conservation. Steady state and homogeneous chemical composition are also 
assumed. The density stratification is set by the momentum equation (the advection term is
not considered) along 
with the mass-loss rate and the wind velocity field (a standard $\beta$-law)
via the equation of continuity. A smooth transition between the wind regime and the pseudo-static
photosphere, which is described by a depth-dependent pressure scale-height, is ensured.

Each model is defined by a set of parameters: effective temperature, surface
gravity and stellar radius (all these quantities are defined at $\tau_\mathrm{Ross}$ =
2/3), the exponent of the wind velocity law ($\beta$), the microturbulence
velocity, the mass-loss rate, the wind terminal velocity and a set of
chemical abundances. In order to reduce the number of parameters to be
explored with our grid of models, we follow the procedure described in 
\citet{urbaneja2005}, with a modification in that the microturbulence 
velocity is fixed to 10 \kms~for the calculation of the atmospheric 
structure. For specific information concerning the model atoms used in our calculations, the reader 
is referred to the last reference.

\subsection{Stellar parameters and surface chemical composition}

\begin{figure}[!]
 \begin{center}
 \includegraphics[width=0.95\textwidth]{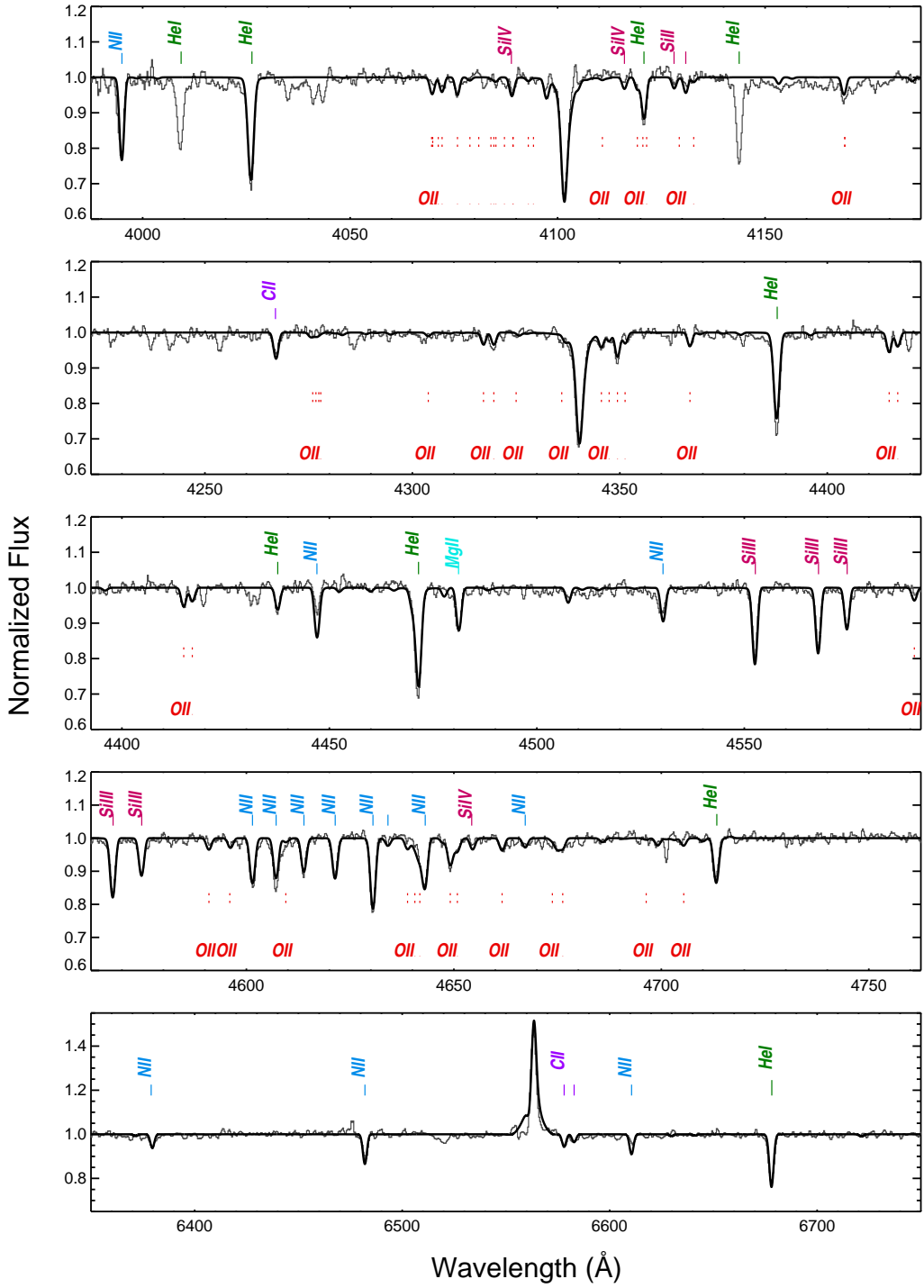} 
 \caption[ ]{Observed ISIS/WHT spectrum (thin line) of UIT\,105 and the synthetic model (thick 
 line) resulting from our analysis. An identification of some prominent features is 
 provided.  \label{fig23}}
 \end{center}
\end{figure}

\begin{figure}[!]
 \begin{center}
 \includegraphics[width=0.8\textwidth]{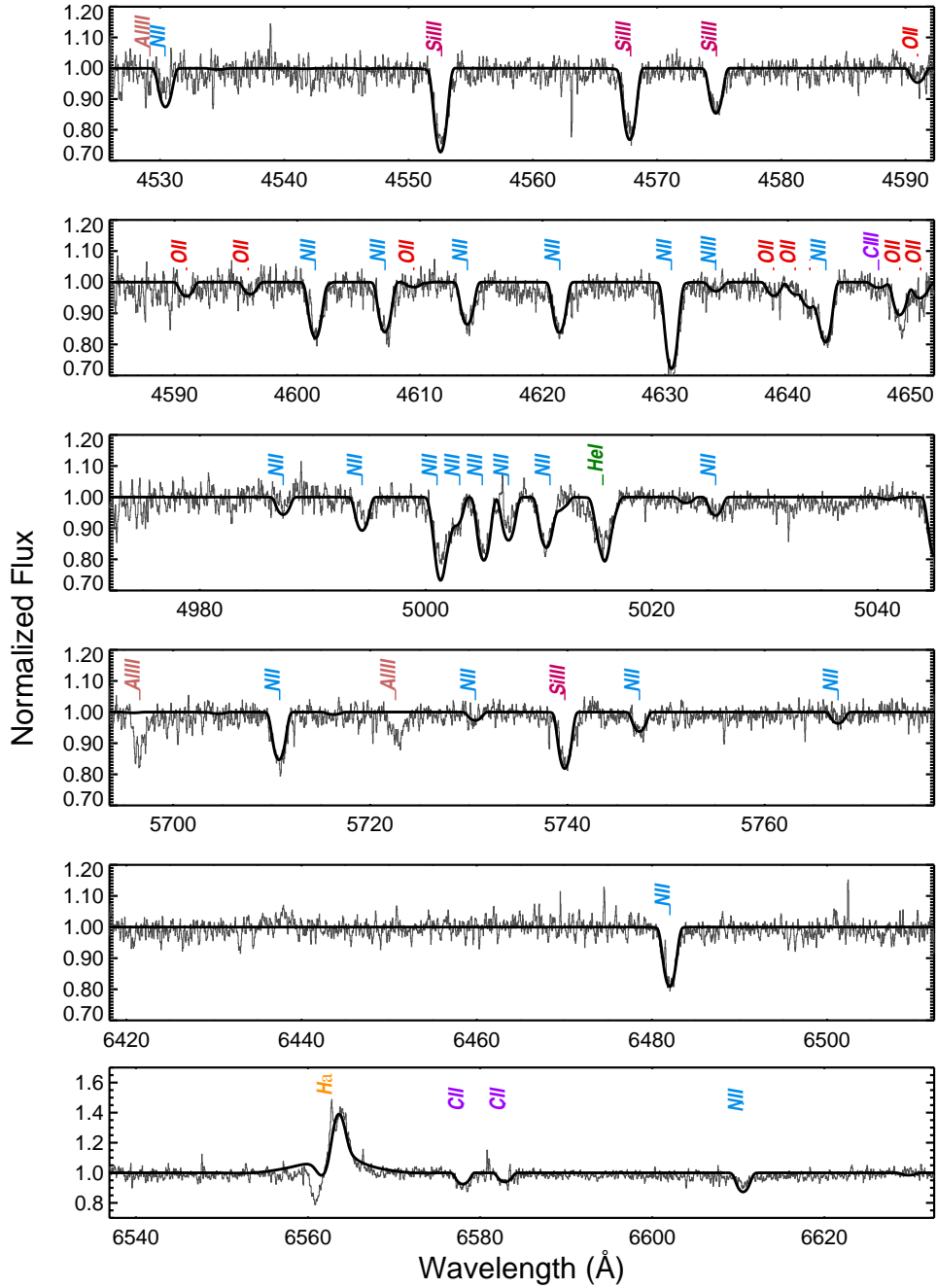}
 \caption[ ]{Observed HIRES/Keck\,1 spectrum (thin line, R$\sim$65000) of UIT\,005 compared with
 our final model. Only selected windows are shown, to illustrate the
excellent agreement between the model based on the lower resolution ISIS/WHT
spectra and the high resolution HIRES spectrum.  \label{fig_hires}}
 \end{center}
\end{figure}

Tab. \ref{u005prop} summarizes the derived properties of UIT\,005, with
Fig.~\ref{fig23} and Fig.~\ref{fig_hires} displaying respectively a comparison of the ISIS
and HIRES data with our final model. The effective temperature is
well constrained by using the \ion{Si}{2} / \ion{Si}{3} ionization balance, and the surface gravity
is obtained from the fits of H$\delta$ and H$\gamma$. The shape of the emission H$\alpha$ profile
allows us to constrain the wind-velocity law ($\beta$), although the profile is not
completely well reproduced, in particular the blue ward absorption. This points to limitations in
some of the assumptions in our modeling technique, in particular related to the density distribution. 
However, 
the derived mass-loss rate is not seriously affected by these possible deficiencies due to the fact 
that the emission part of the profile reacts strongly to changes in this quantity. Also, both 
H$\gamma$ and H$\delta$ profiles appear well reproduced, which reinforces our confidence on the 
global procedure.

In order to constrain the He abundance, we consider the 
\ion{He}{1} lines at 4026, 4713 \AA~(both belonging to the triplet spin system) 
and the one at 4387 \AA~(singlet spin system), while 6678 and 4471 \AA~(singlet and 
triplet systems, respectively), although observed, are not directly
considered. The lines at 4009 (singlet), 4129 (triplet), 4145 and 4437 \AA~(both singlet lines) 
are also not considered because presently our models do not include detailed data for the 
Stark broadening profiles for these lines.

Concerning the N abundance, lines from single ionized N are used to determine its surface 
abundance. These are 3995, 4447, 4601-07-14-21-30 \AA~along with the 
lines present in the red wavelength coverage, namely 6379, 6482 and 6610 \AA.  Other \ion{N}{2} lines 
can be identified at 4035-41-43 \AA, but they involve high energy terms,  
presently above the energy cut-off considered in our \ion{N}{2} model.   

The derived carbon abundance is based on the three available lines:  \ion{C}{2} 4267, 6578-82 \AA. 
It must be noted that we regard our current C model atom as not fully reliable, and thus the quoted C 
abundance has to be considered carefully.

For the determination of the O abundance we consider the single ionized oxygen lines 
located at 4072-76, 4317-19-66, 4414, 4591-96, and 4661 \AA, 
neither of them known to present any blend with any other species in the range of effective
temperatures characteristic of our star. Other \ion{O}{2} lines, not directly used, are present in the
spectrum, such as the lines around $\sim$ 4650 \AA, the weak lines at 4673-6 \AA~and,
particularly, the group of strong lines just redwards of the red wing of H$\gamma$
(located at 4345, 4347, 4349 and 4351 \AA). In general this group of lines forms on the red wing of 
H$\gamma$, whilst due to the narrow H Balmer profiles, these lines appear out of the wing in this case.  

Finally, the Mg abundance is based on the only one feature present in the spectrum, the
\ion{Mg}{2} line at 4481 \AA. \\

\subsection{Error analysis}

Uncertainties affecting the determination of the effective temperature and the (effective)
surface gravity basically depend on the quality of the observed spectrum (i.e. the signal-to-noise
ratio). In our case, Teff can be constrained within a range of $\pm\,$1000 K\,: any larger variation
in this quantity would shift the Si ionization equilibrium in such a way that either \ion{Si}{4} or
\ion{Si}{2} lines would became too strong relative to the observed ones. 

The effective surface gravity
is well determined by simultaneously fitting the H Balmer lines (H$\delta$ and H$\gamma$ in this
case), from which we found an uncertainty of $\pm\,$0.05 dex. To that, we have to add the
contributions coming from the stellar wind parameters. In doing so, the total budget amounts to 
$\pm\,$0.10 dex, taking into account very conservative errors for $\beta$ and $\mathrm{Q^\prime}\,$
(the optical depth invariant, see f.e. \citealt{kudritzki2000}, is defined as  
$\mathrm{Q^\prime}\,=\,\dot{M}\,f_\mathrm{cl}^{1/2}\,\left(R_*\,v_\infty\right)^{-3/2}$~, where 
$f_\mathrm{cl}$ represents the clumping factor\footnote{Presently, clumping is treated under the 
micro-clumping formalism: the wind consists of overdense clumps embedded in a void inter-clump medium; 
$f_\mathrm{cl}$ measures the overdensity in the clumps.}, set to unity in our models).
We would like to stress out that here we are disregarding any possible contribution from the
models themselves, i.e. we consider that our models properly account for the photospheric line
pressure. This assumption is related to the completeness of the line list used in the
blanketing/blocking calculations (the reader is referred to \citealt{puls2005} for a complete 
discussion concerning this point). A second issue is related to the lack of information about the 
Fe abundance. Here we
are considering that the Fe abundance (actually the abundance of any element other than He, C, N, 
O, Mg and Si) scales with the
$\alpha$-elements according to the solar ratio and the global metallicity of the model. This could
have an impact on the effective gravity (again, through the contribution to the photospheric line
pressure). As a sanity check, we have tested that varying the Fe-group element abundances by
$\pm\,$0.2 dex, no noticeable effect is observed on the synthetic H profiles used for the
determination of the surface gravity. 

Finding the stellar radius requires the knowledge of the absolute magnitude, for which
the distance to the star \citep[840 kpc,][]{freedman2001} and photometric data in several 
bands are required. Along with M$_\mathrm{v}$, 
the radius depends also on the SED provided by the models, and in particular on the 
flux in the Johnson V-band. The error in the absolute magnitude is dominated in our particular case by
the uncertainty of the distance modulus to the galaxy, $\Delta\mu\,=\,\pm\,0.10\,$ mag
(see previous reference), since in comparison the adopted error in the apparent
magnitude m$_\mathrm{v}$ is small, $\Delta\mathrm{m}_\mathrm{v}\,=\,\pm\,0.05\,$ mag. This
$\Delta\mathrm{M}_\mathrm{v}\,=\,0.10\,$ mag translates into a contribution of 
$\Delta\,\mathrm{R}(\mu)\,=\,4$
R$_\odot\,$ due to the uncertainty in the distance. We can also estimate the contribution due to the
differences in the theoretical SEDs (basically the uncertainties on Teff) by computing the required
radii to reproduce the absolute magnitude for our limiting models with Teff$\,\pm\,\Delta$Teff,
resulting in $\Delta\mathrm{R}(\mathrm{Teff})\,=\,4\,$R$_\odot\,$. Considering both 
contributions, we estimate an uncertainty in the radius of $\Delta\mathrm{R}\,=\,6\,$R$_\odot\,$.\\

In principle, it would seem feasible to estimate the wind terminal velocity from the H$\alpha$ P Cygni profile present 
in the HIRES spectrum. In doing so, the velocity at 
which the continuum level is recovered corresponds to $\sim$125 \kms. However, the P Cygni profile is not saturated, therefore it 
is unclear whether this is the actual terminal velocity; it could happen that the opacity in the line beyond this point is 
negligible and it is not contributing to the formation of the line. Models computed for a broad range of terminal velocities 
(keeping $\dot{\mathrm{M}}\mathrm{v}_\infty^{-2}$ constant) produced indistinguishable profiles, hence suggesting that the line 
should not be used to estimate this quantity. Without precise information of 
the wind terminal velocity, the goodness of our derived mass-loss rate $\dot{\mathrm{M}}$, is unsure. It is possible to assign
formal errors to $\dot{\mathrm{M}}$ by model comparisons, but it is important to understand that the
derived mass-loss rate is anchored to the adopted wind terminal velocity, in the sense that two models
with different $\dot{\mathrm{M}}$ and v$_\infty$ but keeping 
$\dot{\mathrm{M}}\mathrm{v}_\infty^{-2}$ constant will show very similar H$\alpha$ profiles. 

Recent UV studies of B-type supergiants in different metallicity environments
\citep{bresolin2002, urbaneja2002, evans2004} as well as theoretical investigations 
\citep{puls2000, vink2000, kudritzki2002, krticka2006}, indicate a weak 
metallicity dependence of the terminal velocity. On the other hand, observational studies of Galactic
stars (see \citealt{kudritzki2000} for a review) show a relatively large dispersion in wind terminal 
velocities within a given spectral type.  
For the sake of the analysis, we have adopted a terminal velocity given by
the surface escape velocity and the 
v$_\infty$--v$_\mathrm{esc}$ relationship provided by \citet{kudritzki2000}. The mass-loss rates quoted in 
Tab.\,\ref{u005prop} correspond to the values obtained from the two different H$\alpha$ profiles available, 
and, to avoid any over interpretation, we do not include a precise estimation of
their uncertainties.\\

Finally, at the spectral resolution used in this work, the signal-to-noise ratio is the main
contributor to the uncertainties in the chemical abundances. Other factors could increase the 
line-to-line scatter for each element, like the continuum normalization and the accuracy of the derived 
fundamental stellar parameters. The uncertainties presented in Tab.~\ref{u005prop}, result from the combination (quadratic sum) of two
different contributions. The first one corresponds to the statistical error associated
with the elemental abundances derived from our final model, based on a
line-by-line analysis (see Tab.~\ref{table_ew}). For the second one, we
consider the uncertainties related to the determination of the fundamental
parameters, taking into account the uncertainties affecting the effective temperature, the surface gravity 
(which included the effect of wind parameters), the microturbulence and the uncertainty in the 
helium abundance. These individual contributions are compiled in Tab.
\ref{table_sigma_final}, where the 1$\sigma$ uncertainties quoted in the
last column of the table are obtained as the quadratic sum of the
corresponding contributions.\\

\section{Discussion\label{disc}}

Besides the faint source already discussed in Sect.~\ref{section_photometry}, 
there is no indication in the different spectra that would suggest that 
UIT\,005 is in fact a multiple system. If indeed the star has companions, these are not
contributing at any significant level. There is also no reason to assume
that our object and the nearby faint source are physically related in any way. In the following, we 
consider that UIT\,005 has evolved as a single star, since this is the simplest assumption that can 
be made from the information in hand. Moreover, we adopt the characteristic metallicity of UIT\,005 
as given by the Si and Mg abundances, $\sim$0.7\,Z$_\odot$, i.e. a metallicity intermediate between that of 
the Milky Way and that of the LMC. This value, based on the abundances of two $\alpha$-elements, is consistent with what would be inferred from the C+N+O by mass
relative to the solar value, $\sim$0.8\,\,Z$_\odot$. We note here that this metallicity is higher than the corresponding value for the ionized gas in NGC\,588 when using the gas phase O 
abundance (see below) as a proxy for metallicity, 0.4Z$_\odot$. The population synthesis based study of the ionizing cluster of NGC\,588 by  \citet{jamet2004} indicates an even lower
metallicity, 0.2Z$_\odot$. However, both these results are contradicted by  \citet{bianchi2004}, who find near solar metallicity for UIT\,008, one  of the two known Wolf-Rayet stars in NGC\,588.  \\

\begin{figure}[!]
 \begin{center}
 \includegraphics[]{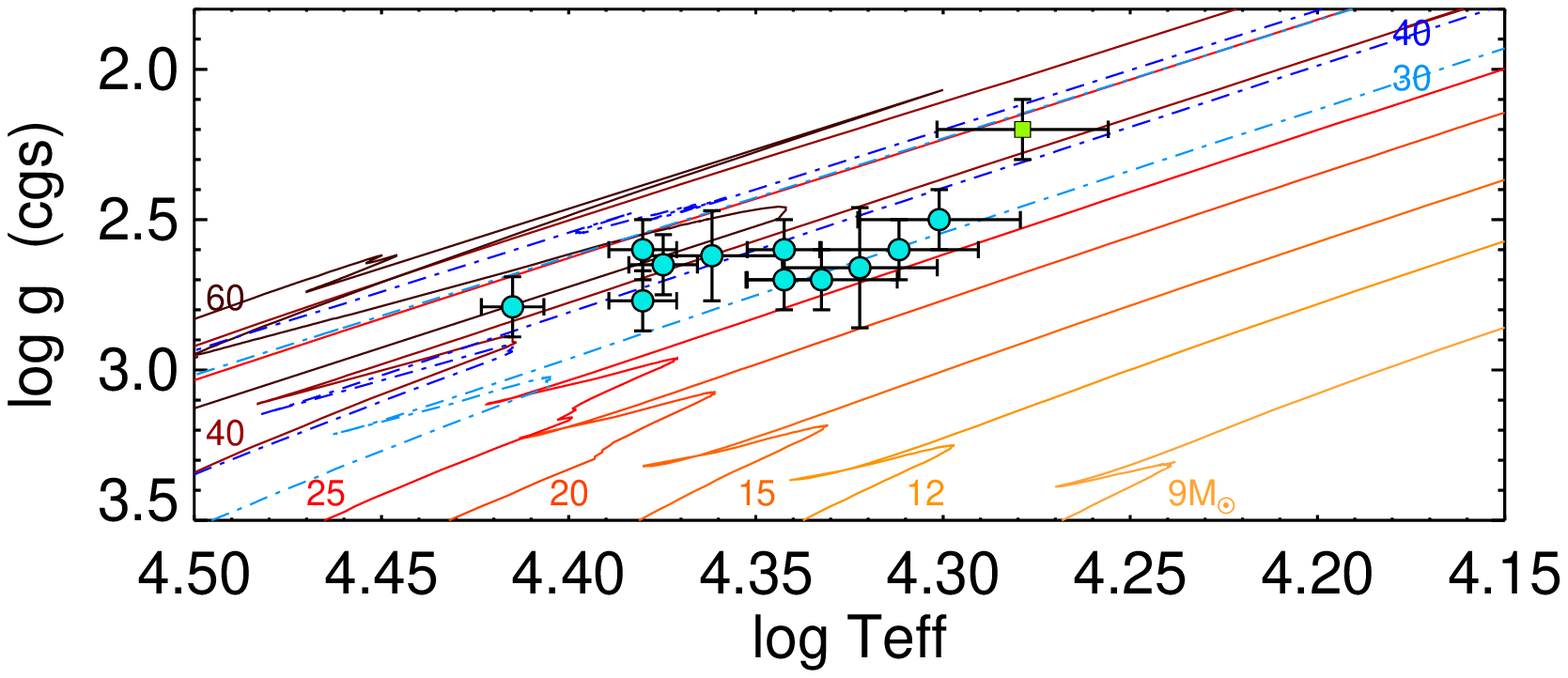}
 \caption[ ]{Derived Teff--log g compared with the predictions of stellar
 evolutionary models accounting for stellar rotation from \cite{meynet2003} for solar metallicity (tracks
 for ZAMS masses 9, 12, 15, 20, 25, 40 and 60 M$_\odot$ are displayed as
 solid lines) and from \cite{meynet2005} for z=0.008 (30 and 
 40 M$_\odot$ tracks are shown as dashed-dotted lines, and are labeled at the upper 
 right corner). The B-type supergiant stars of \citet{urbaneja2005} are shown as circles,
 with UIT\,005 represented by the square. \label{fig5}}
 \end{center}
\end{figure}

\begin{figure}[!]
 \begin{center}
 \includegraphics[]{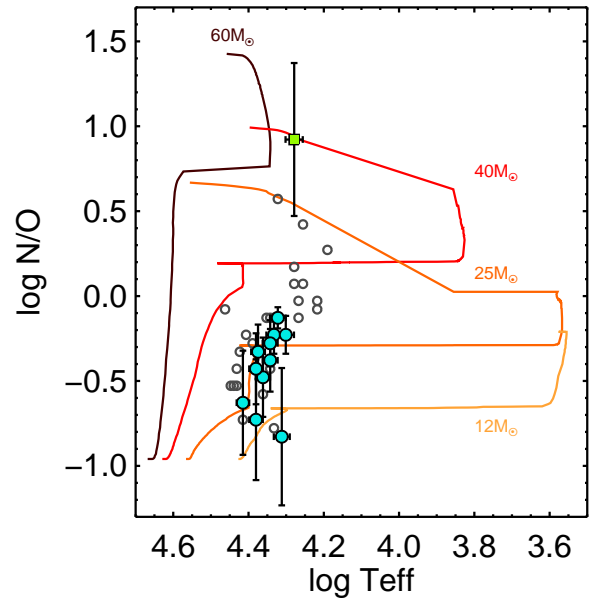}
 \caption[ ]{Surface N/O ratio (by mass) against the effective temperature. Theoretical
 predictions and filled symbols as in previous figures. For reference, the sample of 
 Galactic B-type supergiants studied by \citet{crowther2006} is also shown
 (small symbols without error bars). \label{fig8}}
 \end{center}
\end{figure}

Fig.~\ref{fig5} locates UIT\,005 in a log~Teff--log~g
diagram, together with the sample of M\,33 blue supergiant stars analyzed by 
\citet[][hereinafter UHK05]{urbaneja2005}, and evolutionary
models for Galactic \citep{meynet2003} and LMC \citep{meynet2005} metallicities. The 
position of UIT\,005 is
consistent with models having initial masses between 40 and 60\,M$_\odot$
for both metallicities. As the derived metallicity of UIT\,005 lies between
that of the Galaxy and the LMC, we conclude that its initial mass
also was between these two values. Note that, unlike in the classic Hertzsprung--Russell (H-R) 
diagram (luminosity versus effective temperature), both Teff and log\,g are directly derived from
the spectral analysis, without any assumption regarding the distance to the object. This figure 
also illustrates that UIT\,005 is farther away from the Main Sequence phase than the M\,33 
B-type supergiant stars in the sample studied by UHK05. This point is stressed in Fig.\,\ref{fig8}; our object appears clearly isolated in this N/O--Teff diagram, showing a much 
higher N/O ratio (by mass)  than any of the other M33 objects. The star presents
a N/O$\sim$8, and a N/C$\sim$32 (this last value is highly uncertain because of the issues 
with our C model atom).  In comparison, the object with the higher N/O ratio in the Galactic sample studied by \citet[][empty circles in Fig.~\ref{fig8}]{crowther2006} is the well known blue hypergiant HD\,152236, for which 
these authors find N/O$\,=\,$ 3. As we will discuss later, the apparent agreement with the predictions
by the evolutionary models is misleading.

Even though the abundances are far from the expected values at (CNO) equilibrium, 20.5 and 64 respectively, the very high observed N/O ratio of UIT\,005 is quite remarkable and, at a first glance, 
could be an indication of an advanced evolutionary status (enhanced N abundance). Alternatively, this could 
be a consequence of an initial low O abundance, because of the location of the star in the galaxy\footnote{Normalized distance to the galactic center, $R\,/\,R_{25}$\, = 0.80}  and
its strong radial abundance gradient \citep[UHK05; ][]{u2009}. Or it could result from a combination of 
both.\\  

\begin{figure}[!]
 \begin{center}
 \includegraphics[]{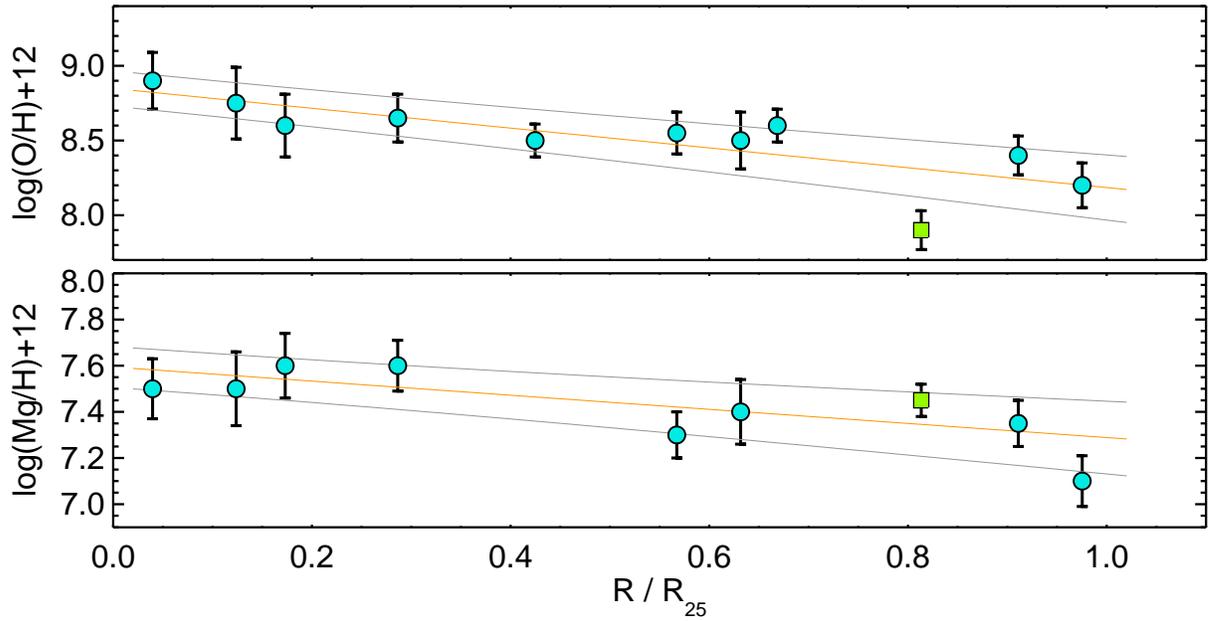}
 \caption[ ]{UIT\,005 oxygen and magnesium abundances in the context of M\,33 radial 
 metallicity gradient. B supergiant stars from UHK05 are shown as circles, 
 with the square representing UIT\,005. The x-axis corresponds to the normalized de-projected 
 distance to the galaxy center, for the distance to M33 and its structural parameters defined in 
 \S\ref {section_intro}. The 1$\sigma$ confidence intervals for the gradients are shown by the 
 thin lines.  \label{fig_o_grad}}
 \end{center}
 \end{figure}

The proposition of an initial low O content can be easily tested by using the information provided by the other
two $\alpha$-elements for which we derived an abundance, Mg and Si. Fig.~\ref{fig_o_grad} displays the O and Mg abundances of UIT\,005 and the sample 
studied by UHK05. In the case of O (Fig.\,\ref{fig_o_grad}, top panel), it seems that UIT\,005 does not follow the gradient defined by the other B-type 
supergiants, whilst this is not the case for Mg (Fig.\,\ref{fig_o_grad}, bottom panel). The same happens if we 
consider Si instead of Mg (not shown). Using the stellar O radial gradient obtained by UHK05, the O abundance 
that would correspond to the galactocentric distance of UIT\,005 is log(O/H)+12\,=\,8.45\,dex. 
Even if one argues that there is no particular reason for the gradient to behave smoothly,  our expectation for 
the O abundance is supported by recent results for the giant \ion{H}{2} region NGC\,588, separated by 
$\sim$30$^{\prime\prime}$ (a de-projected distance of $\sim$130\,pc for the distance to the galaxy adopted in this work) of our 
star, with an O abundance in the range 8.45--8.32\,dex \citep{crockett2006, magrini2007, rosolowsky2008}. It would 
seem then that the present low O content of UIT\,005 is not related to its location in M\,33.  

\begin{figure}[!]
 \begin{center}
 \includegraphics[]{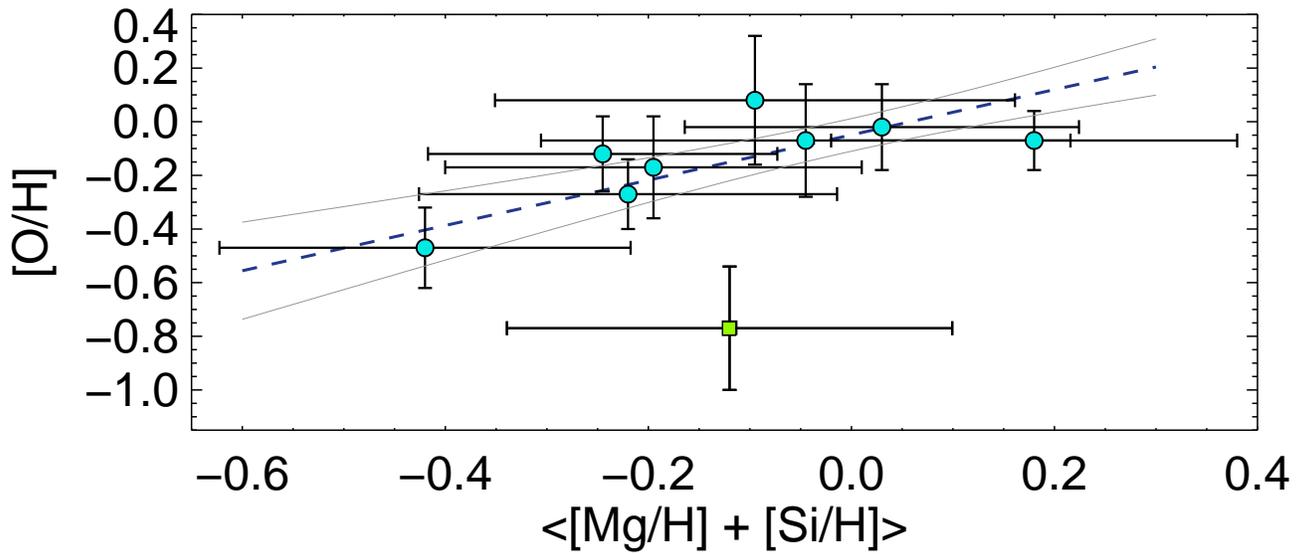}
 \caption[ ]{Oxygen vs magnesium plus silicon abundances, relative to the Sun.
 The square represents the abundances of UIT\,005, and the circles
 correspond to the sample analyzed by UHK05. A linear fit to this sample is
 shown by the dashed line; the 1$\sigma$ confidence intervals are shown by the thin lines. 
 Solar abundances from \citet{asplund2005}, namely: 8.67 dex, 7.56 dex and 7.54 dex for O, Mg and Si,
 respectively. \label{fig_omgsi}}
 \end{center}
\end{figure}

We can further investigate this issue by removing the spatial dependence. Fig.\,\ref{fig_omgsi} presents the 
relation between the derived O abundance and the mean $\alpha$-element abundance defined by Mg and Si. 
Considering the sample of UHK05, it seems clear that there is a relation between the abundances of these three 
elements, that is not followed by the abundances of UIT\,005. Therefore, the relative stellar abundance of O to 
the two other $\alpha$-elements, Mg and Si, in combination with the previous paragraph, strongly support the 
idea of UIT\,005 having suffered significant evolution and presently exposing material processed by the CNO 
cycle.\\

Further constraints can be obtained from the evolutionary lifetimes of
phases consistent with the effective temperature and gravity 
derived for UIT\,005. %%We can rule out the case of an initial
%% 60\,M$_\odot$ for solar metallicity and high initial rotational velocity.
Not all the possible models have the same likelihood, since in some cases
the compatible stages are predicted to be extremely short. In fact, a close 
examination of the lifetimes indicate that for the rotating models these 
are always only a few hundred years, making it extremely improbable that 
UIT\,005 started its life with a high rotational velocity. On the other hand, 
non-rotating models at 50 and 60\,M$_\odot$ spend respectively 6.7 and 
6.2$\times$10$^4$ years in this phase at Galactic metallicity, 
while the 50 M$_\odot$ at LMC metallicity spends 3.8$\times$10$^4$ years,
giving us a chance to observe such an object. Therefore, from the available set of 
evolutionary models, we are left with the non-rotating models at 50\,M$_\odot$ 
(for both metallicities) and 60\,M$_\odot$ (only for the solar metallicity, but note 
that the non-rotating 60\,M$_\odot$ model was not available for LMC metallicity). For 
all three models the evolutionary predicted masses are between 30 and 44\,M$_\odot$, in 
agreement with the spectroscopic mass, within the error bars. At 
the same time, however, the observed surface enrichment is at odds with the ones 
provided by these evolutionary models; whilst N is within the uncertainties
of the observed value, the models predict that the O surface abundance should be almost an 
order of magnitude higher than observed. Since, owing to rotation, one would expect a higher
degree of evolution in N than in O, the observed values would not be explained by models
with higher initial rotational velocity alone. 

Because of the somewhat peculiar derived abundances,  %%the derived parameters place the star close to the Eddington limit, 
one might ask to what extent our results are 
conditioned by the necessary assumption incorporated in the model atmosphere/line formation code used for the analysis. During
the course of our work,  we calculated a number of models with the alternative code CMFGEN \citep{hillier1998}.  These models
were based on our {\sc fastwind} solution for UIT005; i.e. we did not perform an independent analysis using
CMFGEN models in the same way as described in previous sections. We limited the calculations to a small region of the 
parameter space around the values obtained with {\sc fastwind} instead. A discussion of the differences  
between both codes is clearly beyond the scope of our paper; this would require a larger sample of objects, to investigate possible 
systematic differences. Nevertheless, the \textquotedblleft best\textquotedblright~solution  
found with CMFGEN (T$_\mathrm{eff}\sim$18300 K, $log g\sim$2.15 dex) overlaps, within the errors of the analysis, with our parameters. 
Moreover, the chemical abundances derived with both codes are consistent. In particular, the O abundance obtained with 
CMFGEN, 7.85 dex, is in excellent agreement with our {\sc fastwind} derived value of 7.90 dex .\\

Although somehow uncertain due to the lack of precise information about the wind terminal 
velocity, the mass-loss rates derived from the available H$\alpha$ profiles ($\dot{M}\approx1.0-1.4\times10^{-6}$ \mloss) 
are %%% surprisingly
low when compared with the theoretical value predicted by \citet{vink2000, vink2001}, 
$\dot{M}=1.8\times10^{-5}$ \mloss. We have not found firm indications of a structured wind in 
our admittedly limited set of observations. But even if the wind is clumped
to some extent, this will not bring the observed 
and the theoretical value into a better agreement. Rather, the situation would become even worse, 
since a clumped wind would act in the opposite way, implying a smaller {\em true} mass-loss
(our derived mass-loss rates would be upper limits in that case). This 
is a significant point of disagreement with the evolutionary models, since these apply the theoretical
predictions by \citet{vink2000, vink2001} to prescribe the mass-loss rates during the blue 
supergiant phases. This discrepancy between theoretical predictions and 
H$\alpha$ and radio continuum derived mass-loss rates for mid-B types has been previously
reported for SMC \citep{trundle2005} and Galactic \citep{crowther2006,
markova2008, benaglia2008} supergiants. We refer the reader to \citet{markova2008}, where this 
dilemma has been discussed  at some length. In fact, in the case of UIT\,005, the high theoretical 
mass-loss rate is not consistent either with the observed spectral energy distribution (see below).

The available H$\alpha$ profiles indicate a variation of $\sim$25\% in the 
mass-loss rate from the 2001 to the 2003 observations. At the same time, the few He and metal lines 
available in the same wavelength range suggest that this change does not affect the structure 
of the star significantly. Changes in H$\alpha$ profiles at this level have been 
previously reported for Galactic early B-type supergiants
(\citealt{rivinius1997}; \citealt{crowther2006}).\\

\begin{figure}[!]
 \begin{center}
 \includegraphics[]{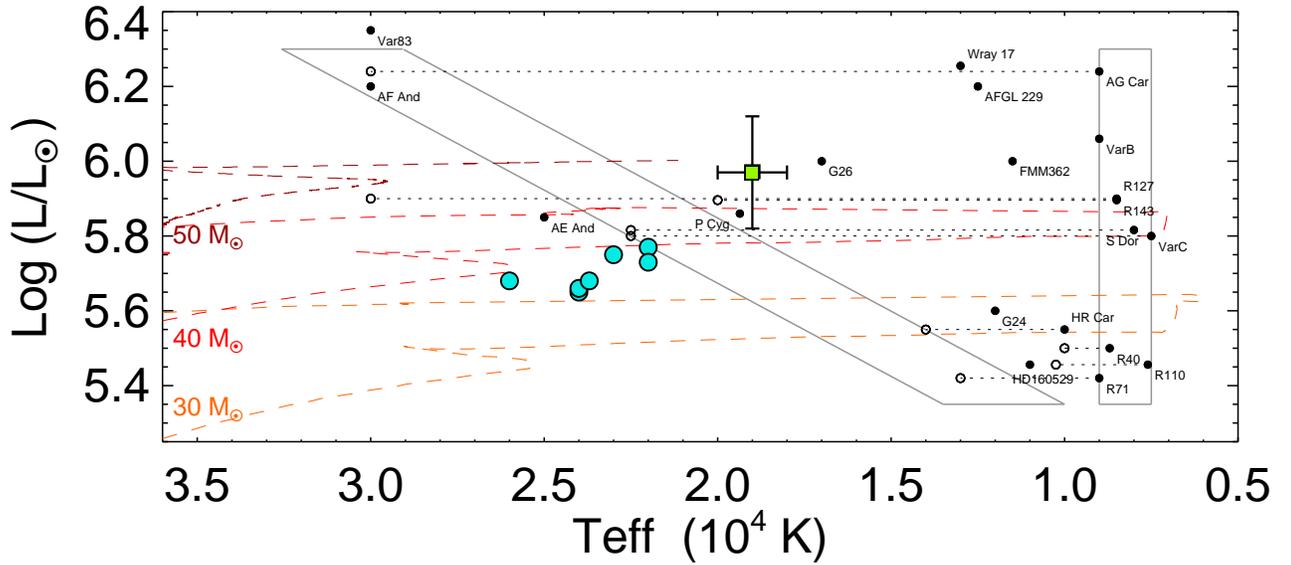}
 \caption[ ]{Hertzsprung-Russel diagram. The square represents UIT\,005, whilst the big dots locate the sample
 of early B-type supergiants studied by UHK05. The approximated location of the S Dor 
 instability strip and the maximum light strip are also indicated.  Known LBVs and LBV candidates are 
 included in this plot (small symbols). When the information is available, the two different states of a 
 given object are connected by dots.
 Evolutionary tracks for Z\,=\,0.008 from the Geneva group are displayed, for
 different initial masses and an initial equatorial rotation of 300~\kms. This figure has 
 been partially adapted from \citet{smith2004}.  \label{fig_lbv_dhr}}
 \end{center}
\end{figure}

\begin{figure}[!]
  \begin{center}
  \includegraphics[]{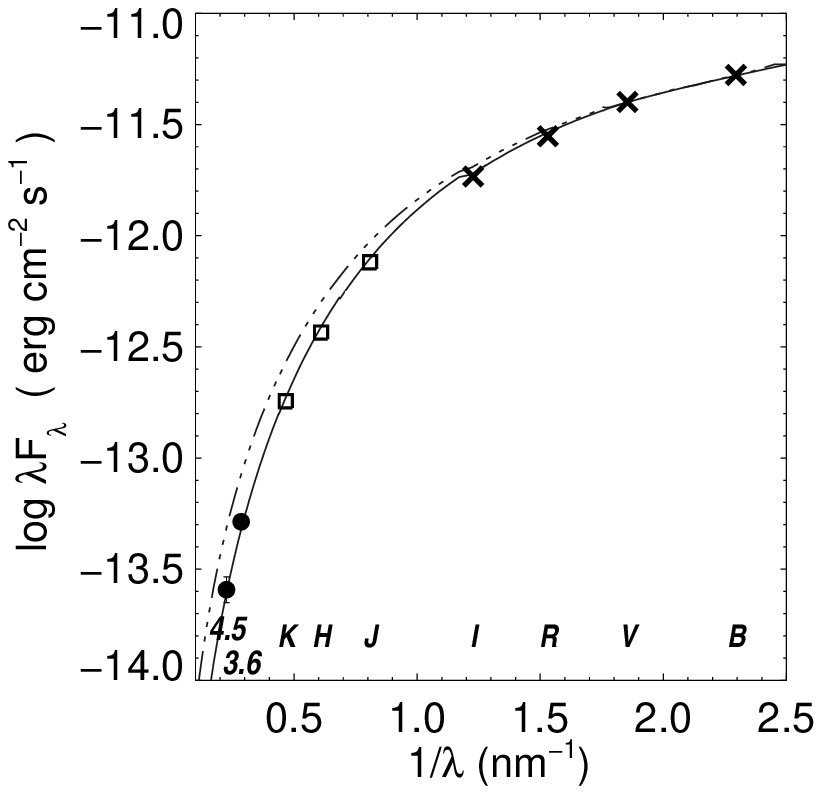} %%% BW figure
  \caption[ ]{UIT\,005's spectral energy distribution. Comparison of the 
   theoretical continuum and photometric measurements
   (symbols) in different bands: B-, V-, R- and I-band photometry from 
   \citet[][crosses]{massey2006}; J-, H- and K-band
   photometry from \citet[][squares]{cioni2009}, and 
   Spitzer IRAC 3.6 and 4.5\,$\mu$m 
   fluxes from \citet[][circles]{mcquinn2007}. 
   The theoretical continua predicted by {\sc fastwind} (solid line--best model,
   dash-dotted line--enhanced $\dot{M}$, see text for explanation) 
   has been reddened following the extinction curve by \citet{cardelli1989}, 
   assuming A$_\mathrm{v}$/E(B-V)\,=\,3.1. 
  \label{fig_ext}}
  \end{center}
\end{figure}

UIT\,005 resides in a region of the H-R diagram (Fig.\,\ref{fig_lbv_dhr}) populated 
by extreme blue objects such as LBVs. In fact, if the distance to M\,33 is larger than assumed 
here, as recently derived from several 
stellar indicators \citep[see][and references therein]{bonanos2006, u2009}, then UIT\,005 could 
be as luminous as $\log\left(\mathrm{L}/\mathrm{L}_\odot\right) \sim 6.1\,$dex, well in the 
regime of the most luminous LBVs. Whether our star is a regular blue object or a dormant LBV is 
unclear. The observed spectral 
energy distribution (SED) from 0.44 to 4.5\,$\mu$m (Fig.\,\ref{fig_ext}) is very well matched by the 
synthetic SED of our final model (Fig.\,\ref{fig_ext}, solid line), even though those photometric 
measurements were acquired in different epochs. The synthetic SED has been reddened assuming a 
total-to-selective extinction ratio R$_\mathrm{v}\,=\,3.1$ and by adopting the extiction curve by 
\citet{cardelli1989}. We note that the
selection of the extinction curve has little or no effect on the considered photometric bands, and the
excellent agreement between observed and theoretical SEDs would not be affected. We have included a 
second model (dash-dotted line in Fig.\,\ref{fig_ext}) computed for an enhanced mass-loss rate, set to  
 the theoretical prediction discussed above. As previously indicated, the observed SED for wavelengths 
longward the I-band is clearly not consistent with this high mass-loss rate.  

With respect to our final model, the good agreement in the infrared bands, and in particular the no 
detections at 8.0 and 24\,$\mu$m (this last from inspection of archival Spitzer data), would indicate the 
lack of a significant amount of  hot dust around the star, suggesting that UIT\,005 is not surrounded  %% hot and cold
by a nebula. Whilst the presence of such nebula is a common characteristic for many LBVs and
it would support the LBV nature of UIT\,005, there are also cases of confirmed LBVs (for example 
R110, Sk\,-69$^{\circ}$\,142a, and R85 in the LMC, \citealt{bonanos2009}) where this nebula 
is absent.\\   

\begin{figure}[!]
  \begin{center}
   \includegraphics[]{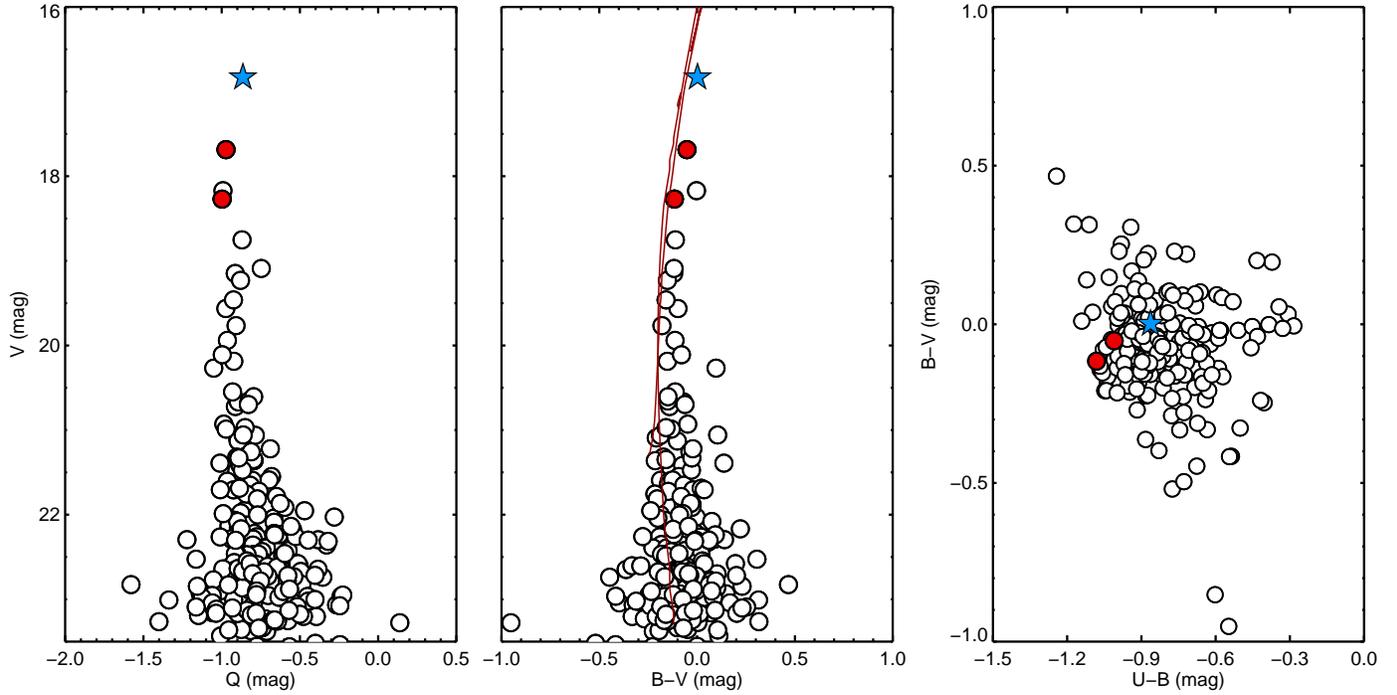} 
  \caption[ ]{Color-magnitude and color-color diagrams constructed from the WFPC2/HST data, covering
  a field of 2.7$\times$2.7 arcmin$^2$ approximately centered at the location of NGC\,588. The 
  {\em reddening free color} $Q$ is defined as 
  $Q\,=\,\left(U-B\right)\,-\,0.72\times\left(B-V\right)$. The two known Wolf-Rayet stars
  in NGC\,588 are identified by filled symbols, with the star marking UIT\,005. For reference, we 
  also show an isochrone for a population $\sim$5 Myrs old. \label{fig_cdm}}
  \end{center}
\end{figure}

The nature of this intriguing object is
further complicated by its apparent isolation,
as illustrated by Fig.~\ref{fig_massey2006} where we see that
it lies well beyond the periphery of its
nearest cluster, NGC\,588.  Is it possible
that this cluster was the original birth-place
of UIT\,005? In Fig.~\ref{fig_cdm} we show the color-magnitude
diagrams (CMD) for all stars in the field (aproximately covering 
2.7$\times$2.7 arcmin$^2$). It is known that NGC588 contains two
WN stars \citep{drissen2008} so it is potentially both young enough and massive
enough to have produced a supergiant such
as UIT\,005. As Fig.~\ref{fig_cdm} demonstrates, UIT005 fits
nicely into the CMD of NGC\,588 as one of the most
evolved of its potential members. %%%(What are the
%%%masses of the WN stars likely to be?)
Furthermore, assuming a main-sequence lifetime
of around 5 Myr (consistent with the age
of 4.2 Myr estimated by \citealt{jamet2004} and \citealt{ubeda2009},  
for a somewhat lower metallicity), it would require a runaway velocity of
around 40-50 \kms~to reach its current distance
from the cluster, well within the range of
velocities of known Galactic runaway O-type stars
\citep{mcswain2007}. While not conclusive, for example there
are no radial velocity measurements for stars
in NGC\,588, the evidence above provides
interesting circumstantial support for the idea
that UIT\,005 is a runaway star from NGC\,588, with the
ejection mechanism being either dynamical
interaction in the cluster, or a SN kick (from a yet more massive companion). 
This second scenario would require a much higher 
runaway velocity, since the SN progenitor would need some time to evolve. Whilst this 
seems to favor the dynamical interaction as the ejection method, very little is know
about the cluster itself, hence this scenario cannot be investigated further at this point.

\section{Conclusions\label{conc}}

Our detailed quantitative analysis of UIT\,005 shows that this massive and luminous 
blue supergiant 
(B2-2.5Ia$^+$, $\log \mathrm{L}/\mathrm{L}_\odot\sim5.9\,\mathrm{dex}$, and M$\sim$50\,M$_\odot$)
has a surface abundance pattern that presents peculiarities,
in particular when considering its location in the disk of
the M\,33. From the point of view of the $\alpha$-elements for which we have
information, Mg and Si abundances seem to follow the radial 
behavior (as traced by other B-type supergiants), while its O abundance appears to be
depleted by a factor $\sim\,$3.5 with respect to the expected local value. The 
$\alpha$/O ratios are hardly understandable as a product of previous
stellar generations at this particular location in the galaxy, moreover if 
we incorporate the derived high N/O ($\sim$8, by mass) and the He abundance. All 
together, and
under the assumption that the star is evolving as a single object, 
the most likely explanation is that the observed chemical pattern is a 
consequence of the chemical evolution of the star, that is showing an advance
stage of CNO processing in its surface. This picture is supported by the predictions 
of recent evolutionary  models, although we must point out that some discrepancies are 
still present, such as the degree of chemical processing predicted by the models.

The derived parameters place the star in a region of the H-R where LBVs are usually found. The 
observations do not support its nature as an LBV, nor they rule out the possibility of an LBV in 
a dormant state. UIT\,005 could very well represent a transitory stage in the evolution of 
the most massive stars or even the precursor of an LBV. Subsequent monitoring of this 
object would help us to improve our knowledge of massive stars, bridging the gap between 
regular and more exotic blue supergiants.\\

\acknowledgments 

We would like to thank J. Puls, R.-P. Kudritzki and Z. Gazak for providing
comments to an early version of this manuscript. D. Bersier and M.\,R. Cioni 
are kindly acknowledged for providing us with their photometric measurements. The
authors wish to thank D.\,J. Hillier for making his suite of codes CMFGEN/CMF\_FLUX
available to us. The anonymous referee is acknowledged for constructive comments.

MAU acknowledges support by NSF under grant AST-10088798. AH and DJL acknowledge 
support by the Spanish MICINN under
grant AYA2008-06166-C03-01 and the Consolider-Ingenio 2010 Program grant 
CSD2006-00070: First Science with the GTC (http://www.iac.es/consolider-ingenio-gtc).
AH also acknowledges support by the Spanich MICINN under grant 
AYA2010-21697-C05-04 and from Gobierno de Canarias ProID20100119.

This publication makes use of data products from the Two Micron All
Sky Survey, which is a joint project of the University of Massachusetts
and the Infrared Processing and Analysis Center/California Institute of
Technology, funded by the National Aeronautics and Space Administration
and the National Science Foundation. 

This research has made use of the Keck Observatory Archive (KOA), which is 
operated by the W. M. Keck Observatory and the NASA Exoplanet Science Institute 
(NExScI), under contract with the National Aeronautics and Space 
Administration.

%%%% BIBLIOGRAFIA %%%%%%%

%%%%%%%%%%%%%%%%%%%%%%%%%%%%%%%%%%%%%%%%%%%%%%%%%%%%%%%%%%%
%%%%%%%%%%%%%%%%%%%%%    TABLES   %%%%%%%%%%%%%%%%%%%%%%%%% 
%%%%%%%%%%%%%%%%%%%%%%%%%%%%%%%%%%%%%%%%%%%%%%%%%%%%%%%%%%%

\clearpage
\begin{deluxetable}{l c c c l}
 \tabletypesize{\scriptsize}
 \tablecaption{Photometric data\label{table_phot}}
 \tablewidth{0pt}
 \tablehead{ 
 \colhead{ }                & \colhead{Magnitude}       & \colhead{}           & \colhead{ } \\
 \colhead{Bandpass/Color}   & \colhead{(mag)}           & \colhead{Date}       & \colhead{Comment } 
 }   
 \startdata
 V     &  16.78                & 1993-11-22 & \citet{massey1996}, ground \\	       
 B-V   &   0.03                & 1993-11-22 & \citet{massey1996}, ground \\	      
 U-B   & -0.91                 & 1993-11-22 & \citet{massey1996}, ground \\
 F170W & 15.566$\,\pm\,$0.013  & 1994-10-28 & this work, WFPC2/HST, no transformation \\
 F336W & 15.415$\,\pm\,$0.004  & 1994-10-28 & this work, WFPC2/HST, U$=$15.970$\,\pm\,$0.004 \\
 F439W & 16.813$\,\pm\,$0.004  & 1994-10-28 & this work, WFPC2/HST, B$=$16.833$\,\pm\,$0.004 \\
 F469N & 16.845$\,\pm\,$0.014  & 1994-10-28 & this work, WFPC2/HST, no transformation \\
 F547M & 16.836$\,\pm\,$0.003  & 1994-10-28 & this work, WFPC2/HST, V$=$16.831$\,\pm\,$0.003 \\
 V     &  16.727$\,\pm\,$0.003 & 2001-09-18 & \citet{massey2006}, ground \\
 U-B   &  -1.053$\,\pm\,$0.003 & 2001-09-18 & \citet{massey2006}, ground  \\
 B-V   &   0.010$\,\pm\,$0.003 & 2001-09-18 & \citet{massey2006}, ground  \\
 V-R   &   0.019$\,\pm\,$0.003 & 2001-09-18 & \citet{massey2006}, ground  \\
 R-I   &   0.058$\,\pm\,$0.003 & 2001-09-18 & \citet{massey2006}, ground  \\
 J     &   16.79$\,\pm\,$0.135 & 1997-12-05 & \citet{cutri2003}, 2MASS \\     
 J    &   16.77$\,\pm\,$0.01   & 2005-09-29 -- 2005-12-16 & \citet{cioni2009}, WFCAM/UKIRT \\	
 H    &   16.74$\,\pm\,$0.01   & 2005-09-29 -- 2005-12-16 & \citet{cioni2009}, WFCAM/UKIRT \\	
 Ks   &   16.77$\,\pm\,$0.03   & 2005-09-29 -- 2005-12-16 & \citet{cioni2009}, WFCAM/UKIRT \\	 
 IRAC 3.6$\,\mu$m & 16.637$\,\pm\,$0.078 & 2004-01-09 to 2005-08-25 & \citet{mcquinn2007}, mean values  \\
 IRAC 4.5$\,\mu$m & 16.659$\,\pm\,$0.145 & 2004-01-09 to 2005-08-25 & \citet{mcquinn2007}, mean values \\

\enddata
\end{deluxetable}

\clearpage
\begin{deluxetable}{l c}
 \tabletypesize{\scriptsize}
 \tablecaption{Stellar parameters and chemical abundances \label{u005prop}} 
 \tablewidth{0pt}
 \tablehead{ \colhead{Properties} & \colhead{UIT\,005} }
 \startdata
Alt. ID       & J013242.92+303847.0                                  \\  
SpT           &  B2-2.5\,Ia$^+$	                                    \\ 
T$_\mathrm{eff}$ (kK)             & 19.0$\,\pm\,$1.0                \\
log g (cgs)                       & 2.25$\,\pm\,$0.10               \\
$v_\infty$ (\kms)                 & 450                             \\
$\dot{M}$ ($10^{-6} $ \mloss)     & 1.42--1.00                      \\
R (R$_\odot$)                     & 86.5$\,\pm\,$6.0                \\
$\beta$                           & 2.20$\,\pm\,$0.20               \\
n(He)/n(H)                        & 0.15 	                    \\
$v_{turb}$ (\kms)                 &  10$\,\pm\,$3                   \\
$v\,\sin i$  (\kms)               & $\le$ 50                        \\
A(C)           & 7.40$\,\pm\,$0.20                      \\ 
A(N)           & 8.85$\,\pm\,$0.15                      \\ 
A(O)           & 7.90$\,\pm\,$0.22                      \\ 
A(Mg)          & 7.50$\,\pm\,$0.16                      \\ 
A(Si)          & 7.35$\,\pm\,$0.15                      \\ 
Z ($Z_\odot$)\,\tablenotemark{a} & 0.75                 \\
\,m$_v$   (mag)                  & 16.79$\,\pm\,$0.05   \\
\,B\,-\,V (mag)                  &  0.01$\,\pm\,$0.01   \\
                                 &                      \\
\,E(B\,-\,V) (mag)               &  0.18$\,\pm\,$0.015  \\
\,M$_v$ (mag)                    & -8.33$\,\pm\,$0.11   \\
\,BC (mag)                       & -1.79$\,\pm\,$0.12   \\
\,Log (L/L$_\odot$) (cgs)        &  5.95$\,\pm\,$0.16   \\
\,M$^\mathrm{spec}$ (M$_\odot$)  & 48.6$\,\pm\,$18.0    \\
\enddata
\tablenotetext{a}{The characteristic metallicity Z is defined as the mean of the individual differences
of Mg and Si abudances with respect to their solar values}
\tablecomments{ The mass-loss rates correspond to the values derived from each H$\alpha$ profile.
}
\end{deluxetable}

\clearpage
\begin{deluxetable}{l r r r r }
 \tabletypesize{\scriptsize}
 \tablecaption{Equivalent widths, line-to-line abundances and statistical uncertainty of the
 mean values \label{table_ew}}
 \tablewidth{0pt}
 \tablehead{ 
 \colhead{ }             & \colhead{EW}         & \colhead{ }        & \colhead{A$\left(\mathrm{X}\right)$ } \\ 
 \colhead{Feature}  & \colhead{ (m\AA) } & \colhead{SNR}  & \colhead{ (dex) }  
 }   
 \startdata
 C{\sc ii} 4267 & 104 & 60 & 7.37 \\
 C{\sc ii} 6578 & 154 & 70 & 7.44 \\
 $\sigma$(C)  &        &      & 0.07 \\
 
 N{\sc ii} 3995 & 365 & 34 & 9.06 \\
 N{\sc ii} 4447 & 117 & 57 & 8.25 \\
 N{\sc ii} 4601 & 177 & 65 & 8.75 \\  
  N{\sc ii} 4607 & 173 & 62 & 8.85 \\   
 N{\sc ii} 4614 & 144 & 60 & 8.82 \\
 N{\sc ii} 4621 & 189 & 86 & 8.91 \\
 N{\sc ii} 4630 & 358 & 67 & 9.00 \\
 N{\sc ii} 6380 & 123 & 84 & 8.75 \\
 N{\sc ii} 6482 & 316 & 63 & 8.87 \\
 N{\sc ii} 6610 & 140 & 99 & 8.51 \\
 $\sigma$(N)  &        &      & 0.08 \\
 
 O{\sc ii} 4072 &   48 & 50 & 7.84 \\
 O{\sc ii} 4076 &   87 & 50 & 7.54 \\
 O{\sc ii} 4317 &   32 & 60 & 7.63 \\
 O{\sc ii} 4319 &   93 & 60 & 8.40 \\
 O{\sc ii} 4366 &   42 & 70 & 7.76 \\
 O{\sc ii} 4414 &   81 & 65 & 7.97 \\
 O{\sc ii} 4591 &   22 & 55 & 7.38 \\
 O{\sc ii} 4596 &   22 & 50 & 7.53 \\
 O{\sc ii} 4661 &  100 & 58 & 7.95 \\
 $\sigma$(O)  &        &      & 0.11 \\
 
 Mg{\sc ii} 4481 &  176 & 74 & 7.50  \\
 $\sigma$(Mg)  &        &       & \ldots \\

 Si{\sc iii} 4552 &  272 & 68 & 7.32 \\
 Si{\sc iii} 4567 &  217 & 75 & 7.31 \\
 Si{\sc iii} 4574 &  155 & 80 & 7.38 \\
 $\sigma$(Si)  &        &      & 0.04 \\

 \enddata
\tablecomments{Abundances derived for each line; SNR as measured in the
continuum close to the line.} 
\end{deluxetable}

\clearpage
\begin{deluxetable}{l c c c c c c c }
 \tabletypesize{\scriptsize}
 \tablecaption{Metal abundance uncertainties: contribution from different sources \label{table_sigma_final}}
 \tablewidth{0pt}
 \tablehead{ \colhead{ } & \colhead{+($\Delta\,T,\Delta$\,log g)} & 
    \colhead{-($\Delta\,T,\Delta$\,log g)} & \colhead{+\,$\Delta\,$vturb} 
    & \colhead{-\,$\Delta\,$vturb} & 
    \colhead{-\,$\Delta\,$Y} & \colhead{-\,$\Delta\,$log Q$^\prime$} & \colhead{$\sigma$} \\    
    \colhead{Species} & \colhead{(dex)} & \colhead{(dex)} & \colhead{(dex)} & \colhead{(dex)} & 
    \colhead{(dex)}  & \colhead{(dex)} & \colhead{(dex)} 
 }   
 \startdata
 C   &  0.09 & -0.16 & -0.02 &  0.02 & -0.04 & -0.10 & 0.19 \\
 N   & -0.07 &  0.05 & -0.10 &  0.09 &  0.00 & -0.03 & 0.13 \\
 O   &  0.05 & -0.17 & -0.04 &  0.04 & -0.03 & -0.06 & 0.19 \\
 Mg  & -0.10 &  0.11 & -0.06 &  0.06 &  0.02 & -0.09 & 0.16 \\  
 Si  & -0.03 & -0.14 & -0.03 &  0.04 & -0.02 &  0.02 & 0.15 \\  
 \enddata
 \tablecomments{The last column presents the 1$\sigma$ uncertainties obtained as the 
 quadratic sum of the different contributions. These 1$\sigma$ uncertainties are combined with the line-to-line abundance
 scatter (see previous table) to produce the finally adopted metal abundance uncertainties quoted in Tab.~\ref{u005prop}. } 
\end{deluxetable}

%%%%%%%%%%%%%%%%%%%%%%%%%%%%%%%%%%%%%%%%%%%%%%%%%%%%%%%%%%%
%%%%%%%%%%%%%%%%%%%%%    FIGURES  %%%%%%%%%%%%%%%%%%%%%%%%% 
%%%%%%%%%%%%%%%%%%%%%%%%%%%%%%%%%%%%%%%%%%%%%%%%%%%%%%%%%%%

\clearpage

\clearpage

\clearpage

\clearpage

\clearpage
 
\clearpage

\clearpage

\clearpage

\clearpage

\clearpage

\clearpage

\end{document}